\documentclass[a4paper,10pt]{IEEEtran}

\usepackage{authblk}  
\usepackage{array}
\usepackage{graphicx} 
\usepackage{multirow}  
\usepackage{amsmath}
\usepackage{multirow} 
\usepackage{amssymb}
\DeclareUnicodeCharacter{2212}{-}
\usepackage[utf8]{inputenc} 
\DeclareUnicodeCharacter{0301}{\'{e}} 
\usepackage[T1]{fontenc} 
\usepackage{float}
\usepackage{caption}
\usepackage{booktabs}
\usepackage{array} 
\usepackage{tabularx}
\usepackage{comment}
\usepackage{xcolor}
\usepackage{hyperref}
\usepackage{tikz}

\newcommand{\setCommonTableSettings}{
    \small 
    \renewcommand{\arraystretch}{1.2} 
    \setlength{\tabcolsep}{10pt} 
}

\definecolor{lime}{HTML}{A6CE39}
\DeclareRobustCommand{\orcidicon}{%
    \begin{tikzpicture}
    \draw[lime, fill=lime] (0,0) 
    circle [radius=0.16] 
    node[white] {{\fontfamily{qag}\selectfont \tiny ID}};
    \draw[white, fill=white] (-0.0625,0.095) 
    circle [radius=0.007];
    \end{tikzpicture}
    \hspace{-2mm}
}

\newcommand{\orcidauthorA}{0000-0003-3148-3765} 
\newcommand{\orcidauthorB}{0009-0000-4700-1682} 
\newcommand{\orcidauthorC}{0000-0002-8337-079X}
\newcommand{\orcidauthorD}{0000-0002-9991-6822}
\newcommand{\orcidauthorE}{0000-0001-6731-3755}
\newcommand{\orcidauthorF}{0000-0003-4666-6847}
\newcommand{\orcidauthorG}{0000-0003-4274-8298}

\newcommand{\orcid}[1]{\href{https://orcid.org/#1}{\orcidicon}}

\title{Security and Real-time FPGA integration for Learned Image Compression}

\author[1]{Mazouz Alaa Eddine\orcid{\orcidauthorA}}
\author[2]{Carl De Sousa Trias\orcid{\orcidauthorB}}
\author[1]{Sumanta Chaudhuri\orcid{\orcidauthorC}}
\author[3,1]{Attilio Fiandrotti\orcid{\orcidauthorD}}
\author[1]{Marco Cagnazzo\orcid{\orcidauthorE}}
\author[2]{Mihai Mitrea\orcid{\orcidauthorF}}
\author[1]{Enzo Tartaglione\orcid{\orcidauthorG}}

\affil[1]{LTCI, Télécom Paris, Institut Polytechnique de Paris, France}
\affil[2]{SAMOVAR, Télécom SudParis, Institut Polytechnique de Paris, France}
\affil[3]{Università di Torino, Italy}

\affil[ ]{Email: alaa.mazouz@telecom-paris.fr}

\date{March 2025}

\begin{document}
\maketitle

\begin{abstract}

Learned Image Compression (LIC) has demonstrated superior compression efficiency compared to standardized video codecs. However, achieving both real-time performance and robust security in hardware-based LIC deployments remains a significant challenge. This work addresses these challenges by presenting an integrated workflow and deployment platform for training, securing, and implementing LIC models on hardware.  
To achieve a hardware-efficient LIC model, we employ an iterative pruning and quantization process within a standard end-to-end learning framework. Additionally, we introduce \textbf{Quantization-Aware Watermarking (QAW)}—a novel technique that embeds a watermark directly during quantization via a joint loss function. This ensures model integrity and security without degrading performance. To further enhance protection, the watermarked weights undergo public-key encryption, safeguarding both content and user traceability.  
We evaluate real-time performance, latency, energy consumption, and compression efficiency across multiple FPGA platforms. Results indicate that the watermarking and encryption steps introduce only a minor overhead—\textbf{PSNR decreases by 0.2 dB on average, energy consumption increases by 2\%, and FPS drops by 6\%}—while maintaining real-time constraints and security properties. Furthermore, our approach outperforms existing hardware-based LIC implementations in both FPS and energy efficiency, delivering optimized LIC codecs for HD, FHD, and UHD resolutions. 

\end{abstract}
Keywords— Learned Image Compression, FPGA, Quantization, Real-time, Watermarking, Encryption.
\section{Introduction}

In Learned Image Compression (LIC) \cite{duan_lossy_2023, theis_lossy_2017, beusen_image_2022, yang_lossy_2024}, 
an autoencoder is trained to encode and decode an image while generating a compressed bitstream in the process \cite{balle_end--end_2017, balle_variational_2018}.
At the source, the image is projected into a low-dimensional latent space by the \textit{encoder} model. This representation is then quantized and entropy-coded into a binary bitstream. At the receiver, the bitstream is entropy-decoded, and the \textit{decoder} model projects the representation back to the pixel domain, recovering an approximation of the original image. 
Over the last years, the LIC efficiency has seen consistent advancements, nowadays challenging standardized coding technologies \cite{TMM_lic, TMM_lic2} and consequently, the interest in its realtime hardware-based LIC implementations emerged.


Research on \textit{realtime} LIC for hardware has received lot of attention, tackling challenges such as computational/storage/power efficiency \cite{noauthor_arithmetic_2021,fowers_g_nodate}, cross-platform support \cite{gan_energy-efficient_2016,ODE-ODR}, and compatibility with edge computing devices \cite{choi_trainware_nodate, mazouz_adaptive_2019, mazouz_automated_2021}.
Achieving real-time operations on hardware usually relies a combination of strategies. The model is first pruned to minimize the number of filters or neurons \cite{tan_dropnet_2020, cheng_survey_2023}.
Then, the surviving parameters are quantized \cite{gholami_survey_2022} to cope with the fixed-point computation capabilities of embedded hardware. For example, \cite{jia_fpx-nic_2022} and \cite{sun_real-time_2022} propose a FPGA implementations of LIC models with 8-bit parameter quantization.
The state of the art architecture \cite{sun2024fpga} achieves 37 fps encoding and decoding at 720p resolution on a KU115 FPGA.

Alas, research in \textit{secure} hardware-based LIC has received comparatively less attention, despite its growing importance in the broader landscape of multimedia technologies. The LIC paradigm shifts the value to the methods used to design and train the encoder and/or decoder models that are deployed on user hardware. Without secure mechanisms in place, the unlicensed use of these technologies can disrupt the chain of value that fuels innovation in the multimedia industry. This issue is particularly pressing when considering the deployment of LIC models on hardware. In such scenarios, ensuring that i) unlicensed use of trained models is detectable, and ii) the source of any leaks is traceable becomes not just beneficial but essential.

Neural network watermarking \cite{TMM_wm} consists of hiding
an imperceptible fingerprint within a model parameters at training time, so that the model integrity can be verified later on \cite{ boenisch_systematic_2021}.
Watermarking a LIC models requires that the watermark does not alter the video quality or encoding efficiency and survives potential attacks \cite{TMM_wm_attack}.
Watermarking a model deployed on hardware \cite{WM-FPGA} is even more challenging due to the constraint of hiding the watermark within the few parameters surviving the pruning step.
However, attacks aside, the quantization step alone tends to destroy the watermark that cannot be easily retrieved anymore.
Atop of that, retrieving and verifying the watermark must not compromise realtime operations.
Therefore, securing a hardware based LIC model while preserving its properties is a significant challenge that has received little attention to date and is the object of this work.

This work advances a unified workflow for training a hardware-friendly LIC model, securing it through watermarking and encryption, and finally deploying the secured model on an FPGA.
As a preliminary step, a typical LIC architecture is modified and iteratively pruned so as to reduce its computational complexity.
Next, we introduce Quantization Aware Watermarking (QAW), an iterative process in which the model is watermarked and quantized at the same time by fine-tuning it over a loss function that embeds the watermark into the model while keeping the error on the quantized model at bay.
Security is finally achieved through a Digital Rights Management (DRM) scheme that incorporates watermarking, model encryption, and secure key management.
A different watermark is embedded in each model at quantization time using the client public key.
Subsequently, the watermarked  weights are encrypted to protect them during deployment or storage on the hardware accelerator.  
The client is authenticated using their public-private key-pair on a specific device. This key-pair is also used to exchange the encryption key for the model parameters stored on the device.  
At image compression time, the key is employed to decrypt and load the weights onto the accelerator platform.  
Should the model weights be leaked, whether through reading device memory or side-channel attacks, the source of the leak can be traced back to the client device, thanks to client-specific watermarking.  
If the illicit user attempts to erase the watermark, its removal would jeopardize the image quality, rendering the leaked model unusable
Our scheme relies on the widely used Xilinx Vitis-AI \cite{vitis_ai} framework and is designed to be, for the most part, hardware-agnostic, ensuring compatibility with a wide range of accelerators.
We experiment securing a reference LIC model with the above method and we deploy it over two different FGPA platforms.
Our experiments show that our method can embed a robust watermark that secures the model without noticeably impairing encoding speed, quality or rate.

The outline of this paper is as follows. In Section \ref{sec:background}, we introduce the background required for understanding this work. In Section \ref{sec:proposed}, we describe our pipeline for training a hardware-friendly model that is both realtime capable and secure through the steps of pruning, quantization aware watermarking and encryption. Finally, in Section \ref{sec:experiments} we experimentally validate our method, whereas Sec \ref{sec:conclusions} summarizes the lessons learned and discusses future research directions.
The main contributions of this work are as follows:  

\begin{itemize}  
    \item We propose an integrated workflow that combines iterative pruning, quantization-aware training, and secure watermarking to generate a learned image compression (LIC) model optimized for real-time FPGA deployment.  

    \item We introduce a novel quantization-aware watermarking (QAW) method that embeds the watermark directly during quantization, ensuring security and traceability without degrading compression efficiency or image quality.  

    \item Our optimized models achieve 61.2 FPS at HD (1280×720) resolution and 24.2 FPS at FHD (1920×1080) resolution on the ZCU102 FPGA, surpassing prior FPGA-based LIC implementations in both speed and efficiency.  

    \item Our FPGA-based deployment achieves the lowest reported power consumption of 2.88 J/frame at HD and 5.44 J/frame at FHD, significantly outperforming existing FPGA-based learned compression systems in terms of energy efficiency.  

    \item We design and implement a secure digital rights management (DRM) system that combines public-key encryption and watermarking, ensuring model protection with minimal overhead, introducing only a 4\% increase in energy consumption and a 1.5\% FPS drop at HD.  

    \item We provide a comprehensive analysis of the trade-offs between compression efficiency, security, and hardware resource utilization, demonstrating that our approach enables real-time, secure, and energy-efficient LIC deployment on edge hardware.  
\end{itemize}  

\section{Background}
\label{sec:background}

This section provides the relevant background on LIC and presents some preliminary results that we will use as a baseline throughout this work. Next, we provide a primer on neural network watermarking, highlighting the reasons that make watermarking a hardware LIC model a challenge.

\subsection{Learnable Image Compression}

\begin{figure}[t]
  \centering
  \includegraphics[width=1.0\columnwidth]{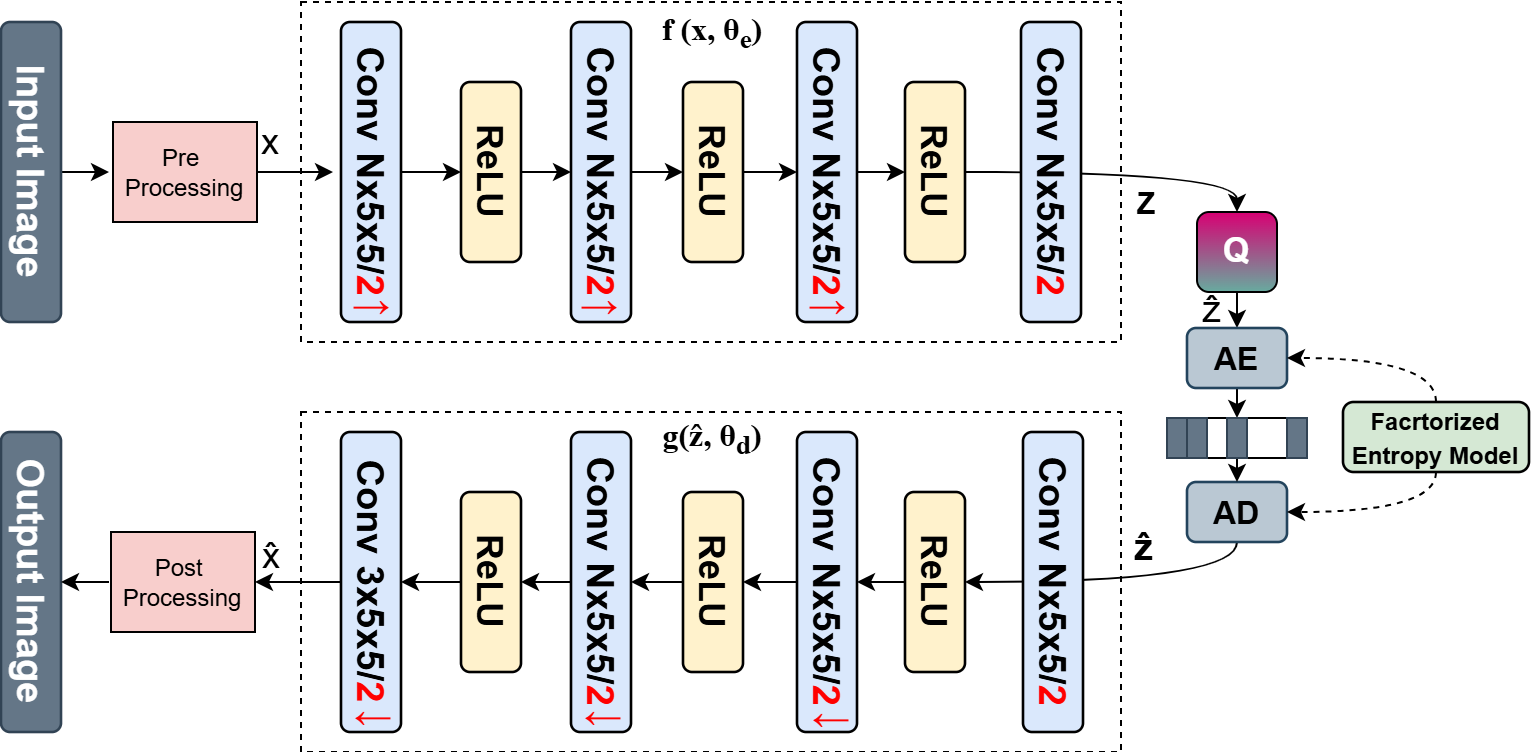}  
  \caption{The reference Learned Image Compression (LIC) model \cite{balle_end--end_2017} we consider in this work, N is set to 192 and the arrows describe the downsampling and upsampling ratio }
  \label{fig:lic2017}
\end{figure}

Fig.\ref{fig:lic2017} exemplifies the seminal LIC architecture \cite{balle_end--end_2017} that we consider in this work.
Borrowing from the original notation, the encoder \( f(\mathbf{x}; \theta_e) \) is composed of a stack of convolution-activate-pool layers that map the input image \( \mathbf{x} \) to a low-dimensional latent representation \( \mathbf{z} \). This latent representation is then quantized into the discrete-valued latent representation \( \hat{\mathbf{z}} \).  
Since the stepwise operator is not differentiable,
uniform noise is introduced in the range [-0.5, 0.5] to emulate the quantization noise \cite{jia_fpx-nic_2022}. This ensures that the decoder \( g(\hat{\mathbf{z}}; \theta_d) \) learns to reconstruct the image from the quantized latent representation available during inference.  
The quantized latent representation is then entropy-coded via Arithmetic Encoding (AE), generating the compressed bitstream. On the receiver side, Arithmetic Decoding (AD) retrieves the latent representation, which is subsequently passed through the decoder to recover an approximate version \( \hat{\mathbf{x}} \) of the input image.

While models such as \cite{balle_variational_2018, minnen_joint_2018, zou2022devil} offer improved RD performance, this work focuses on assessing the feasibility of a real-time yet secure hardware LIC implementation.
By leveraging the simpler architecture of \cite{balle_end--end_2017}, our goal is to demonstrate that meaningful compression can be achieved within the practical limits imposed by hardware constraints. This trade-off between algorithmic complexity and hardware feasibility highlights the need to bridge the gap between theoretical advancements and deployable solutions. Such efforts are crucial not only for establishing the viability of real-time LIC implementations but also for positioning Learned Image Compression as a competitive and practical alternative to existing, widely used codecs.

The above model is trained to minimize the loss function 

\[
L = R + \lambda D \tag{1}
\]
\noindent
where \(R\) is the rate (i.e., the number of bits) of the entropy-coded latent representation \(\hat{Z}\), while \(D\) is the distortion as the  
Mean Squared Error between the original \(I\) and reconstructed  \(\hat{I}\) images.
We point out that such distortion naturally arises in lossy LIC models due to projecting \(I\) to a low-dimensional latent representation and its subsequent quantization (entropy coding is a reversible function and introduces no distortion). Finally, \(\lambda\) is the hyperparameter controlling the trade-off between image quality and the rate of the compressed image. 

Large values of \(\lambda\) imply minimizing distortion (i.e., better video quality) at the expense of an increased image rate. Conversely, small values of \(\lambda\) result in more aggressive quantization, yielding a more compact image representation at the expense of quality. 

\subsection{Neural Network Watermarking}

Watermarking provides a family of methodological and applicative tools making it possible for some metadata (also referred to as mark or watermark) to be imperceptibly and persistently inserted into some original multimedia content \cite{cox2007digital}, according to a secret key. By subsequently recovering these metadata from potentially modified versions of the watermarked content, various applications related to content source identification and integrity verification can be served. 
This simplest watermarking setup brings to light the four main properties among which a practical watermarking solution should find a trade-off:
\begin{itemize}
    \item \textbf{data payload} represents the size of the watermark: the quantity of information than can be inserted and detected.
    \item  \textbf{imperceptibility} refers to preserving the quality of the original content. 
    \item \textbf{robustness} refers to the ability to recover the mark from the content subjected to additional operations (attacks). 
    \item \textbf{secret key} considers the amount of information that should be kept secret.
\end{itemize}

Neural network watermarking emerged in 2017 with the goal of assessing integrity of a neural network \cite{uchida_embedding_2017} and  consists in embedding the watermark within the model parameters \cite{desousa25watermas} .
Inherited from the multimedia watermarking, imperceptibility and robustness properties should be carefully reconsidered.  First, the watermark is not inserted in a multimedia content but in the model itself. Hence, the imperceptibility properties should be transposed to the inference produced by the watermarked model. Moreover, this insertion could occur through training making the mark intrinsically linked to the watermarked content. Finally the inserted watermark could be either retrieved from the parameter of the model \cite{uchida_embedding_2017,desousa25watermas} (white-box case or in the inference of the model \cite{adi2018turning,zhang2018protecting} (black-box case). Second, the robustness now considered NN specific  modifications like fine-tuning (e.g., targeting the performance improvement when slightly changing the task) or pruning (e.g., for fitting the constraints of some embedded devices). 

While most of the watermarking literature focuses on classification tasks, recent works demonstrate that white-box methods can extend beyond this domain \cite{sousa_trias_hitchhikers_2023,trias_find_2024}. Various computer vision models have been successfully watermarked against a wide range of attacks \cite{desousa24perm,desousa25watermas}. However, only \cite{WM-FPGA} applies watermarking at the hardware level, yet it still faces challenges in bridging the gap between algorithmic design and efficient hardware implementation while balancing security, latency, and video quality. 
Notably, no prior work addresses the challenge of watermarking a real quantized Learned Image Compression (LIC) model. This study aims to explore the difficulties of LIC neural network watermarking, posing the following key questions: \textbf{Can a watermark be inserted into a quantized model? Can this insertion be achieved without significant impact on latency and energy or degrading image quality?}

\section{Proposed Method}
\label{sec:proposed}

\begin{figure}[htbp]
  \centering
  \includegraphics[width=0.8\columnwidth]{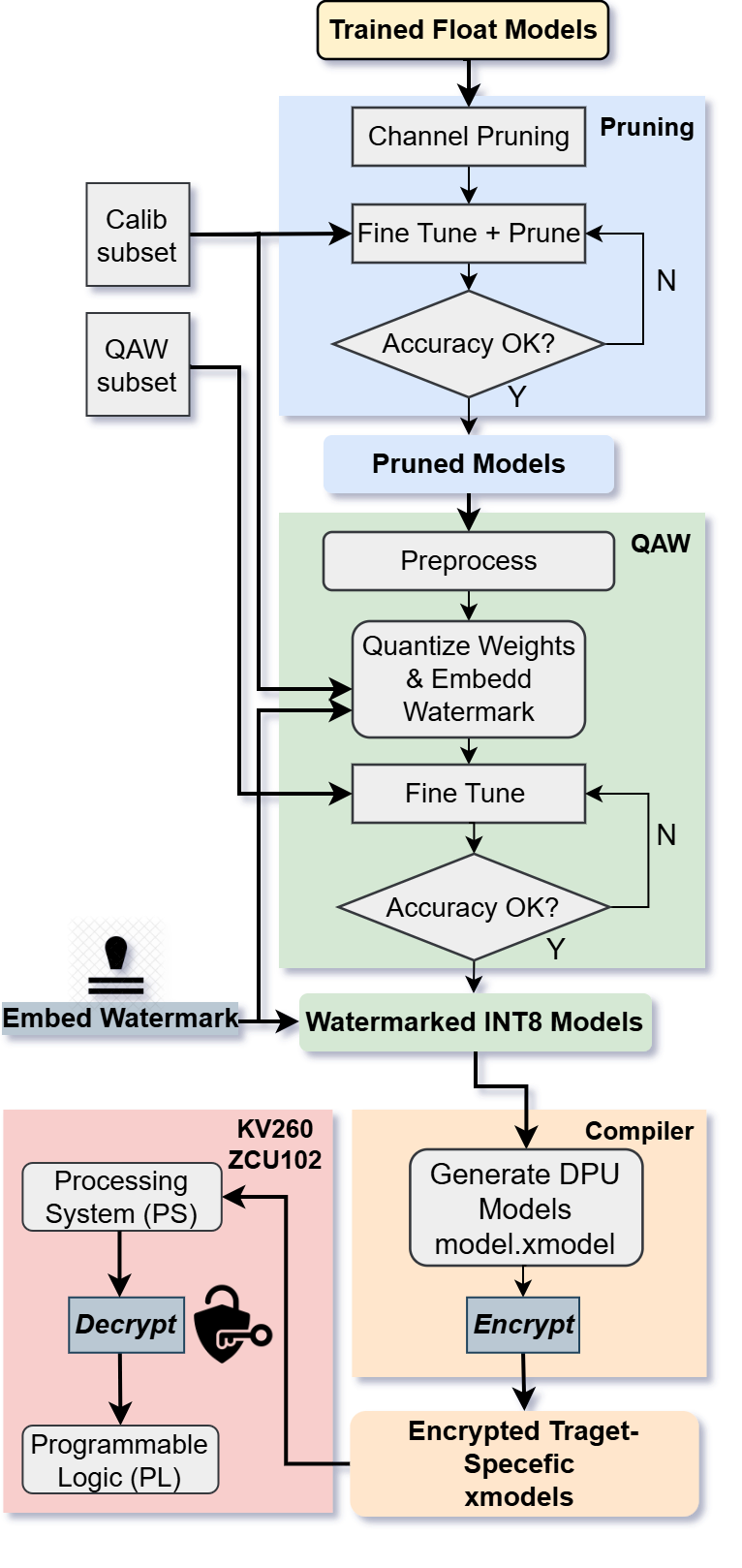}  
  \caption{The proposed workflow for training, optimizing, watermarking, and deploying LIC models on hardware is comprehensive yet abstracts hardware-specific compilation through Xilinx VITIS-AI APIs, ensuring accessibility for non-hardware experts.}
  \label{fig:main_diagram}
\end{figure}

Fig.\ref{fig:main_diagram} details each step of our workflow for securing and deploying on FPGA a LIC model capable of realtime operations.
In existing literature, a (pre-)trained model (cf. previous section) is preliminarily pruned and then watermarking is applied, yet the following quantization may erase the watermark.
With Quantization-
Aware Watermarking (QAW), the watermark is embedded in the model while the latter is quantized by minimizing an ad-hoc loss function.
The workflow takes full advantage of Xilinx’s VITIS-AI API \cite{vitis_ai}, a Python-based tool for compiling and deploying deep neural models on Xilinx boards generating target-specific instructions in the form of xmodels. The compiled model is encrypted using the Advanced Encryption Standard (AES) \cite{rachh_efficient_2012} and saved in the target FPGA external memory.

\subsection{Channel Pruning}
\label{sec:pruning}
Neural networks need to be over-parameterized to be successfully trained, so we prune the model at training time to learn a sparse topology. Due to the convolutional nature of the considered LIC model, we adopt a channel-wise pruning \cite{huang_coarse-grained_2021} approach where entire filters are pruned from the model. While pruning individual parameters may achieve better sparsity, channel pruning is more hardware-friendly and can be implemented with most inference architectures, most importantly with Xilinx Vitis-AI architectures \cite{gao_dpacs_2023}.
Channel pruning practically reduces latency since hardware is not designed to deal with parameter-level sparsity as follows. The original, unpruned, model is given as the input for the optimizer for a first pruning and fine-tuning iteration.
The fine-tuned model obtained at the last iteration becomes the new baseline and is again pruned and fine-tuned. This prune and fine-tune process is repeated until the target sparsity is reached: we set the per-iteration pruning ratio to 0.1 with an increment of 0.1 over 3 iterations, resulting in a target pruning ratio of 30\%, depending on performance. We found that iterative pruning yields a better sparsity-performance tradeoff than single-pass pruning. In the pruned model, many parameters are set to zero during training to induce sparsity. Once pruning is complete, convolutional filters with all-zero weights are removed from the model to reduce computational complexity. If all filters in a layer are pruned, the corresponding feature maps are eliminated. This process ensures that the model's architecture is both sparse and efficient, with unnecessary components entirely removed.  The pruning ratio affects the computational complexity as FLOPs (pruned model) = (1 – ratio) × FLOPs (original model).

\subsection{Quantization Aware Watermarking}
\label{sec:QAW}

In this section we first describe the process of quantizing the weights of a (pruned) model.
Next, we describe our procedure for embedding a watermark in the model while quantizing it.

\subsubsection{Weight quantization}

In order to meet the computational capabilities of FPGAs, the weights (and activations) that have survived pruning must be quantized from 32-bit floating-point to 8-bit integers (INT8)
For weights lower than 1, a scaling factor is applied to normalize the range before converting to INT8 \cite{quant2021white}. This ensures that the quantization range fully utilizes available bit space without truncation errors. 
Minimum and maximum values are extracted from the different model layers on a \textit{calibration set}, allowing the quantizer to derive the optimal scale and the zero point from the float model.
Finally, to achieve symmetric quantization, the zero point is set to zero to enable saturation for quantization and reduce the quantization difference. The quantization process for the weights is defined as follows:
\[
\hat{\theta} = \text{clip}\left( \text{round} \left( \frac{\theta}{s} \right), -127, 127 \right) \cdot s \tag{2}
\]
where \( \hat{\theta} \) represents the quantized weights, \( \theta \) are the original floating-point weights, and \( s \) is the per-layer learned scaling factor. Clipping ensures that the quantized values are within the range of [-127, 127], which is the valid range for symmetric INT8 values. The scaling factor \( s \) is learned during the training process.
Simulated quantization operations are introduced on both weights and activations, instead of using full precision floating-point values; such operations mimic the effects of quantization by reducing the precision of the weights and activations to the target bit-width (8-bit in this work). Figure~\ref{fig:QAT_diagram} shows this process in comparison to standard floating-point data-flow. This simulates the quantization process that will occur during inference on hardware with limited precision. During each forward pass through the network, operations are performed using the quantized weights and activations. These quantized values are then used for computations. During fine-tuning, the gradients are computed with respect to the quantized parameters. This involves propagating gradients through the quantization operations, which are non-differentiable. To address this, a Straight-Through Estimator (STE) \cite{jacob_quantization_2018} is used to approximate gradients through quantization:
\[
\text{STE}(\hat{\theta}) = \theta + (\hat{\theta} - \theta) \quad \tag{3} 
\]
This allows the gradients to be propagated through the quantization layers during training, making it possible to perform backpropagation in the presence of non-differentiable quantization operations.

\subsubsection{Inserting the Watermarking}
\label{sec:proposed_watermarking}

Unlike traditional methods where watermarking and quantization are treated as separate steps, we jointly optimize both during a finetuning process over an ad-hoc loss function.

\begin{figure}[htbp]
  \centering
  \includegraphics[width=0.8\columnwidth]{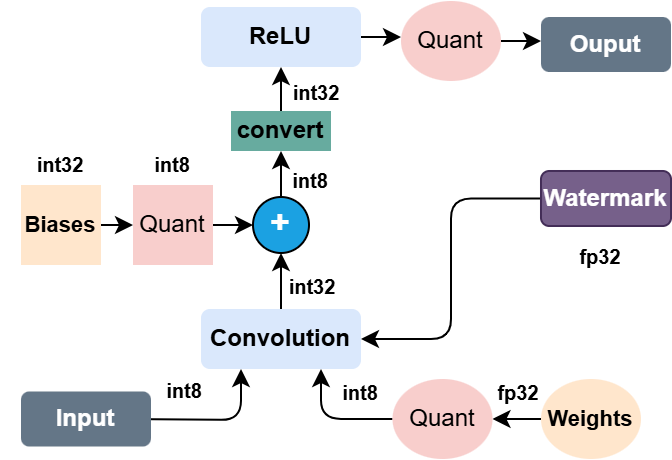}  
  \caption{Quantization Aware Training process dataflow}
  \label{fig:QAT_diagram}
\end{figure}

In this section, we introduce our novel approach to watermarking called Quantization Aware Watermarking.

To ensure that the embedded watermark remains intact after quantization, we integrate watermarking directly into the Quantization-Aware Training (QAT) process using the technique from \cite{uchida_embedding_2017}. Rather than treating watermark embedding and quantization as separate tasks, we jointly optimize both during training. This allows the model to simultaneously account for both quantization errors and watermark embedding, enabling them to be handled together in a single learning phase. Using fake quantizers, we simulate the quantization effects during training while keeping the weights in the floating-point domain \cite{quant2021white} as we embed the watermark, as seen in Fig.\ref{fig:QAT_diagram}. Our approach eliminates the need for separate training stages—first training a quantized model and then fine-tuning for watermark robustness—thereby reducing computational overhead and preventing performance degradation from handling these tasks independently.

The total loss function for this process is defined as:

\[
L_{\text{QAW}} = R + \lambda D + \beta E \tag{4}
\]

Here, \( R + \lambda D\) represents the classic rate-distortion cost function minimized at training time.
The loss term \( E \) ensures the watermark is recoverable after quantization, with embedding quality controlled by the factor \( \beta \). The factor \( \beta \) balances the trade-off between watermark robustness and compression quality to regulate the watermark bit error rate (BER) while maintaining overall model performance. The value of \( \lambda \) is consistent with the value used in floating-point model training, ensuring that rate-distortion performance remains the primary focus.
The watermark loss term \( E \) is defined as
\[
E = \left\| W - X^T \cdot \text{vec}(\hat{\theta}) \right\|^2 \tag{5}
\]
\noindent
where \( W \) denotes the binary watermark, consisting of \( M \) bits, and \( X \) represents a randomly generated matrix of size \( N \times M \) (the key). The term \( \hat{\theta} \) refers to the floating-point weights used during training, with fake quantization ensuring the watermark is embedded before real quantization. Finally, \( \text{vec}(\hat{\theta}) \) is the flattened version of the weights, treated as if quantized but remaining in the floating-point domain for training.
\\
The loss term \( E \) is optimized to minimize the difference between the watermark \( W \) and the projection of the flattened watermarked weights onto the key matrix \( X \), with \( \beta \) controlling the trade-off between watermark robustness and model compression quality. 
\\
By jointly training for both watermark robustness and quantization, the model learns to embed the watermark in the floating-point domain while simultaneously adjusting for the effects of quantization. This ensures that the watermark remains intact even after real quantization is applied and eliminates the need for post-training watermark insertion or additional fine-tuning. This joint approach not only reduces errors introduced by quantization noise but also saves both time and computational resources by avoiding separate fine-tuning steps.

\subsection{Hardware Model Compilation and Deployment}

\begin{figure}[htbp]
  \centering
  \includegraphics[width=0.8\columnwidth]{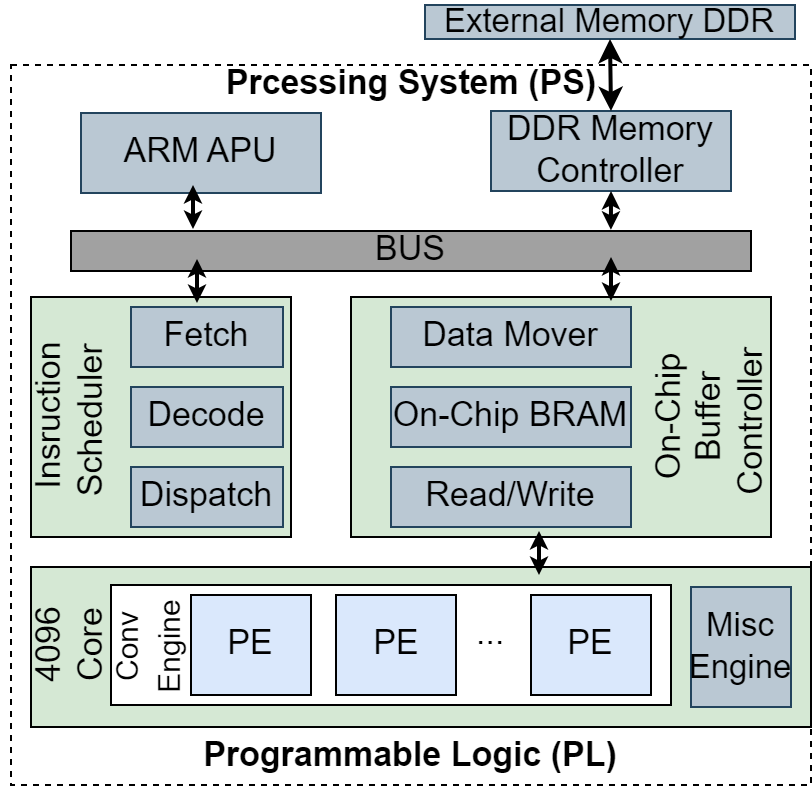}  
  \caption{Hardware architecture of the overall Deep learning Processing Unit}
  \label{fig:DPU_diagram}
\end{figure}

The secured model is compiled into target-specific DPU instructions in the form of an xmodel. The XIR compiler constructs an internal computation graph as an intermediate representation (IR), it then performs multiple optimizations, such as computation node fusion, and efficient instruction scheduling by exploiting inherent parallelism or exploiting data reuse on the specific target. This unified IR is broken down into subgraphs based on the corresponding control flow, for our models a single DPU subgraph is compiled. Finally, an instruction stream is generated for each subgraph and everything is serialized into a target-specific, compiled xmodel file. This xmodel is encrypted and then deployed on the FPGA using the runtime API through the embedded processor.
We deploy the models on the Zynq-UltraScale+ MPSoC DPU architecture -DPUCZDX8L, seen Fig.\ref{fig:DPU_diagram}. The nomenclature refers to the target FPGA, quantization bit-width and design objective, with L being low latency.

The DPU provides multiple configurations to trade off PL resources for performance. We utilize the B4096 core configuration for all compilations, meaning we deploy the engine with the highest IO parallelism (8 pixels,16 input channels, and 16 output channels) thus yielding the highest throughput with a peak of 18×16×16×2=4096 MAC/cycle. A single DPU core is deployed on the KV260 and three on the ZCU102.

\subsection{Secure Deployment and Digital Rights Management}

\begin{figure}[htbp]
  \centering
  \includegraphics[width=1\columnwidth]{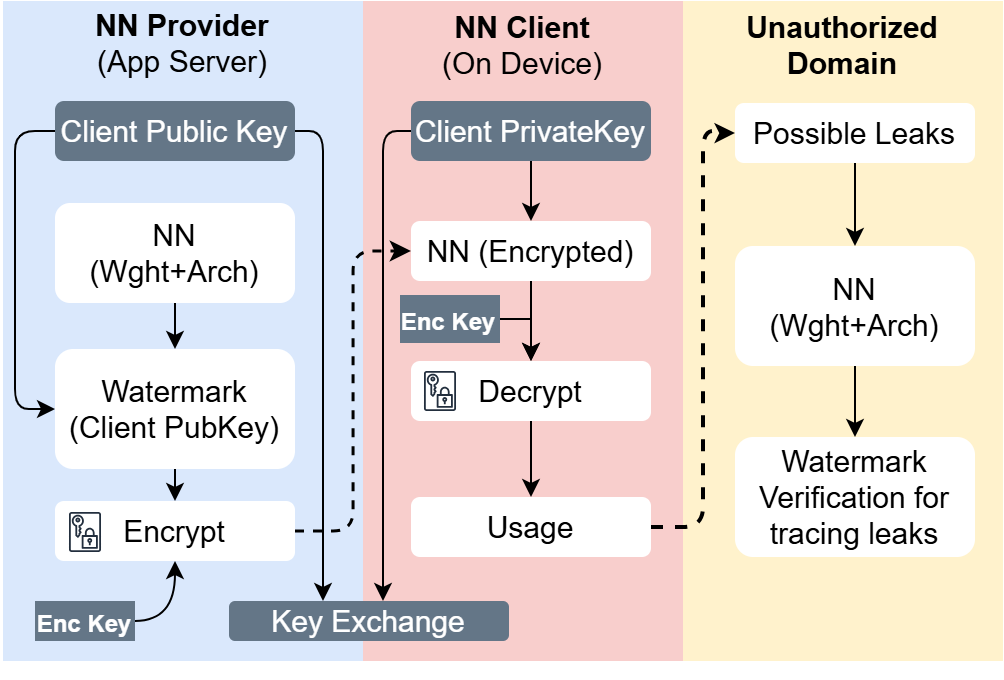}  
  \caption{The proposed DRM scheme}
  \label{fig:DRM_diagram}
\end{figure}

The proposed DRM scheme for the compression platform and application is seen in Fig.\ref{fig:DRM_diagram}. This system is similar to the Widevine \cite{widevine_architecture_2017} DRM model commonly used for multimedia content. The provider of the compression application embeds watermarks into the neural network weights for each client based on their public key. This facilitates tracing the source of a leak and claiming ownership later on. Subsequently, the neural network weights are encrypted by the provider to safeguard them during transmission and storage. The user is authenticated using their public-private key pair on a specific device. This key pair is also utilized for exchanging the encryption key for the model parameters, which are then stored on the device. During the operation of the compression app, this key is employed to decrypt and load the weights onto the accelerator platform.

In detail, the watermarked models are compiled into hardware models, which are then encrypted using AES. We chose AES for its proven security, efficiency, and widespread adoption. As a NIST-approved standard, AES ensures robust data protection and regulatory compliance. Optimized for embedded devices, it offers flexible key sizes to meet system requirements, making it the ideal choice for safeguarding the LIC model parameters and inference data.
We use a 256-bit secret key to safeguard against unauthorized access to the models’ parameters and architecture and disallow tampering. The encrypted models are saved on the target board external memory and accessed only during inference, decryption occurs on the PS and the decrypted DPU model is passed to the PL using the runtime API, the compression models are not consumable without access to a private key.

\section{Experimental Evaluation}
\label{sec:experiments}

In this section, we train baseline float LIC models then evaluate the contribution of each and every step in the workflow proposed in the previous section in terms of compression efficiency, computational complexity and security of the LIC model, comparing with state of the art references where appropriate.


\subsection{Training the Model}
\label{sec:reference_results}

\begin{figure}[t]
  \centering
  \includegraphics[width=1\columnwidth]{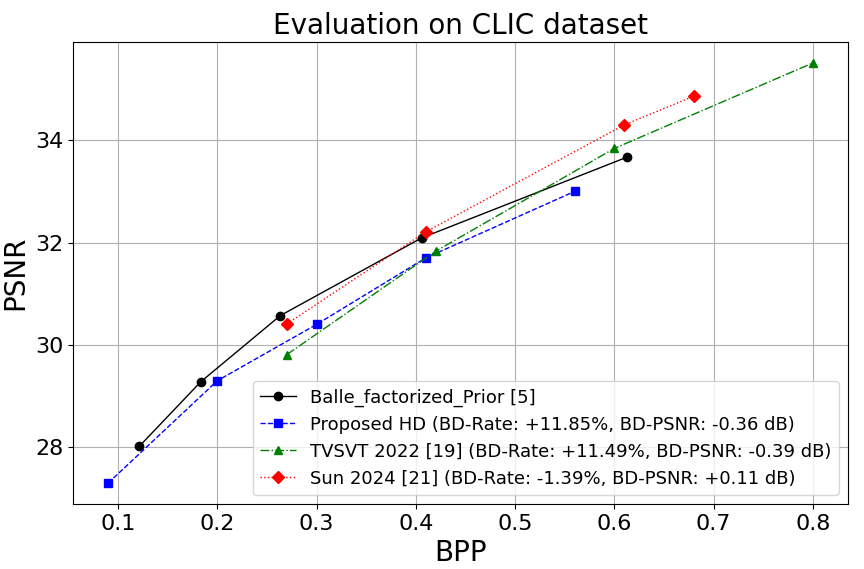}  
  \caption{comparison with reference GDN-factorized model \cite{balle_end--end_2017}}
  \label{fig:HDvsBalle}
\end{figure}

We preliminarily train the LIC model \cite{balle_end--end_2017} from scratch to validate our experimental setup and establish baseline metrics for compression efficiency and latency. Following standard practice, we train five different instances of the model for various values of \(\lambda\) (0.001, 0.005, 0.01, 0.05, 0.1) to cover a sufficiently wide range of rate-distortion (RD) trade-offs. Specifically, we consider a hardware-friendly implementation and training procedure of the model, as shown in Fig.~\ref{fig:lic2017}, which is inspired by related work such as \cite{jia_fpx-nic_2022}. In this approach, the Generalized Divisive Normalization (GDN) layer activation is replaced with a simpler ReLU activation to reduce latency and streamline the implementation for real-time operations.
While \cite{sun2024fpga} achieves higher system-level performance on the same LIC model, that is at the cost of a complex  hardware design space exploration that is out of the scope of this work.
The model is trained using the same randomized train/validate/test dataset split of the combined CLIC, DIV2K, and a subset of the Open Images datasets.
Fig. \ref{fig:HDvsBalle} illustrates the encoding efficiency of our baseline, hardware-friendly implementation of \cite{balle_end--end_2017} compared to the original model. The model presented in \cite{sun2024fpga} is based on the computationally heavier architecture from \cite{balle_variational_2018}, which incorporates an additional hyperprior network instead of a simple factorized entropy encoder. This modification leads to a noticeable improvement in rate-distortion (RD) performance, albeit at the cost of system-level performance. The observed efficiency drop due to the use of a simpler ReLU activation function in place of the GDN is consistent with findings in similar literature. We hypothesize that a hardware-optimized implementation of the GDN could potentially preserve encoding efficiency while minimizing latency impact; however, exploring this is beyond the scope of the present work.


Table~\ref{tab:model_performance_cpu} presents the encode-and-decode latency for different image sizes in a manner consistent with related work \cite{jia_fpx-nic_2022}, measured on a A100-PCIE-40GB GPU.
Following the approach in \cite{jia_fpx-nic_2022}, we generate HD (1280x720), FHD (1920x1080), and UHD (3840x2160) images by repeatedly tiling images from the aforementioned datasets until the target resolution is achieved. 
To account for the memory constraints typical of embedded platforms, the image is processed as partially overlapping 256x256 patches from the tiled images, as in related literature.
Each patch is independently encoded and decoded, and the decoded patches are finally reassembled into the whole picture.
The table shows that the reference LIC model cannot achieve realtime video coding despite our optimizations, emphasizing the need for hardware-tuned LIC models especially for high-resolution contents.

\begin{table}[htbp]
\centering
\caption{Model-specific performance on a single A100-PCIE-40GB GPU. Up/down arrows indicate higher/lower is better.}
\small
\renewcommand{\arraystretch}{1.2}
\setlength{\tabcolsep}{6pt}
\begin{tabular}{l c c c}
\hline
 & \textbf{HD} & \textbf{FHD} & \textbf{UHD} \\ 
\hline
\textbf{GFLOPS} $\downarrow$ & 17.74 & 131.35 & 231.78 \\ 
\textbf{Latency (s)} $\downarrow$ & 0.35 & 0.79 & 3.16 \\ 
\textbf{Avg PSNR (dB) $\uparrow$} & 33.95 & 33.83 & 32.91 \\ 
\textbf{Avg MS-SSIM $\uparrow$} & 0.982 & 0.978 & 0.967 \\ 
\hline
\end{tabular}
\label{tab:model_performance_cpu}
\end{table}

\subsection{Pruning the Model}

\begin{table}[h]
\centering
\setCommonTableSettings 
\caption{Effect of channel-wise pruning compared to the trained LIC model}
\begin{tabular}{p{3.5cm} c c c} 
\hline
\textbf{} & \textbf{HD} & \textbf{FHD} & \textbf{UHD} \\ 
\hline
\textbf{$\Delta$ Model size ↓} & -33.2\% & -27.1\% & -27.1\% \\ 
\textbf{$\Delta$ Inference latency ↓} & -39.0\% & -24.3\% & -29.4\% \\ 
\textbf{$\Delta$ FLOPS ↓} & -30.4\% & -19.6\% & -24.4\% \\ 
\textbf{$\Delta$ Avg PSNR (dB) ↑} & -1.25 & -1.21 & -0.41 \\ 
\textbf{$\Delta$ Avg MS-SSIM ↑} & -6.2\% & -7.7\% & -7.1\% \\ 
\hline
\end{tabular}
\label{tab:pruning_floats}
\end{table}

Next, we assess the effect of pruning the models above following the channel-wise pruning method  described in Sec.~\ref{sec:pruning}.
Table~\ref{tab:pruning_floats} shows that the pruned model has up to nearly 50\% smaller size in terms of number of non-zero parameters, and that is why pruning is so crucial for deployment on hardware with limited resources. Regarding latency, pruning reduces inference time by 29 to 39\%, improving real-time performance, though this comes with a minor trade-off in RD performance. Furthermore, pruning leads to a reduction in FLOPS by up to 30.4\%, which not only accelerates inference but also decreases energy consumption on battery-powered devices. These gains come with tolerable loss in image quality, as indicated by slight decreases in PSNR (-0.1 dB for UHD ) and MS-SSIM, which are generally acceptable for real-time applications where realtime processing is prioritized. 
\\
On the other hand, the table shows that pruning reduces the room available for hiding a watermark in the model parameters by nearly 50\%, complicating the process of securing a model.
These pruned models will be used in the rest of this work as a baseline to assess the effects of the scheme proposed in the next section.

\subsection{ Jointly Watermarking and Quantization the Model }

In this section, we experiment with hiding a watermark within the pruned model weights during the quantization step with our novel QAW strategy as described in Sec.~\ref{sec:QAW}.
Prior to training, we extract the minimum and maximum values from various model layers. These values allow the quantizer to derive the optimal scale and zero-point for the floating-point model, based on the calibration subset we provide. The calibration set for all quantization experiments consists of 1000 randomly selected images from CLIC \cite{toderici_wenzhe_2020}, DIV2K \cite{agustsson_ntire_2017}, and Open Images \cite{kuznetsova_open_2020}, which are used exclusively for the forward pass.

For the actual quantization training, we run 1000 iterations using a randomly selected subset of 5000 images from CLIC, DIV2K, and Open Images. Symmetric quantization is employed with the zero-point set to zero and saturation enabled, minimizing quantization error. We set  \(\lambda\) to the same values used during the floating-point model training, allowing us to explore different rate-distortion (RD) trade-offs (\(0.001, 0.005, 0.01, 0.05, 0.1\)). As for the watermark embedding loss function variable \(\beta\), we explore a range of values based on the chosen \(\lambda\) and target layer in order to find the optimal \(\lambda\) - \(\beta\) pair for each model. To achieve this, we adaptively modify \(\beta\) during QAW to ensure that the watermark is fully embedded by the end of the joint-training, in the event of failure to embed, \(\beta\) is raised, and if embedding is achieved quickly  \(\beta\) is lowered. 
\begin{table}[htbp]
\centering
\setCommonTableSettings
\caption{Performance metrics for pre-watermarked model.}
\begin{tabular}{l c c c}
\hline
\textbf{Lambda} & \textbf{PSNR} & \textbf{MSSIM } & \textbf{BPP} \\
\hline
0.1  & 38.73 & 0.992 & 1.567 \\
0.05  & 37.72 & 0.9899 & 1.205 \\
0.01   & 33.57 & 0.975 & 0.440 \\
0.005   & 31.81 & 0.9627 & 0.261 \\
0.001   & 28.12 & 0.921 & 0.092 \\
\hline
\end{tabular}
\label{tab:noWM_reference}
\end{table}

Once QAW is done, we evaluate the impact of the introduced watermarking on the model by measuring the BD metrics using the float-pruned model as a reference, following the joint QAW training scheme. Table \ref{tab:noWM_reference} presents the PSNR, MSSIM, and BPP values that serve as the baseline for our analysis. To assess the effect of watermarking, we experiment with embedding the watermark across all four layers of the decoder.

We evaluate the robustness of the watermark in terms of C-BER \cite{sousa_trias_hitchhikers_2023}, which quantifies the differences between the embedded and extracted watermarks, expressed in terms of the bit error rate (BER), as follows:
\[
\text{C-BER} = (1 - \text{BER}) \times 100.
\]
\noindent In this case, a higher C-BER value indicates better robustness of the watermark, as it reflects a smaller error between the embedded and extracted watermarks, signifying a more reliable watermark.
The watermark detection is then performed by projecting the watermarked layer onto the secret key, a process executed on the FPGA processing system using the runtime API.

The experimental results are presented in Tables \ref{tab:layer1WM}, \ref{tab:layer2WM}, \ref{tab:layer3WM}, and \ref{tab:layer4WM}. These tables display the PSNR, MSSIM, and BPP performance of the five trained models with the watermark embedded in different layers.It is observed that embedding the watermark in the final layer is particularly challenging, a phenomenon that has been well-documented in the neural network watermarking literature \cite{sousa_trias_hitchhikers_2023}. This difficulty stems from the final layer's heightened sensitivity to perturbations in the network parameters, particularly the weights and activations. Unlike earlier layers, the final layer is responsible for generating the output of the model, where even small changes to the parameters can significantly affect the model's predictions. The final layer typically operates in a more linear regime compared to earlier layers, where activations are often saturated or subject to nonlinearities. Therefore, modifications introduced by watermark embedding in the final layer tend to have a disproportionately large effect on the model's performance, as they directly influence the output logits.
\begin{table}[htbp]
\centering
\setCommonTableSettings
\caption{Performance metrics for Layer 4 with watermarking.}
\begin{tabular}{l c c c}
\hline
\textbf{Lambda / Beta} & \textbf{PSNR} & \textbf{MSSIM} & \textbf{BPP} \\
\hline
0.1 / 0.1 & 38.31 & 0.988 & 1.681 \\
0.05 / 0.1 & 37.31 & 0.981 & 1.275 \\
0.01 / 0.05 & 33.22 & 0.970 & 0.470 \\
0.005 / 0.02 & 31.51 & 0.960 & 0.290 \\
0.001 / 0.01 & 27.86 & 0.910 & 0.096 \\
\hline
\end{tabular}
\label{tab:layer4WM}
\end{table}

\begin{table}[htbp]
\centering
\setCommonTableSettings
\caption{Performance metrics for Layer 3 with watermarking.}
\begin{tabular}{l c c c}
\hline
\textbf{Lambda / Beta} & \textbf{PSNR} & \textbf{MSSIM} & \textbf{BPP} \\
\hline
0.1 / 0.1 & 38.67 & 0.9924 & 1.640 \\
0.05 / 0.1 & 37.51 & 0.989 & 1.244 \\
0.01 / 0.05 & 33.54 & 0.975 & 0.480 \\
0.005 / 0.02 & 31.79 & 0.964 & 0.280 \\
0.001 / 0.01 & 28.09 & 0.920 & 0.0944 \\
\hline
\end{tabular}
\label{tab:layer3WM}
\end{table}

\begin{table}[htbp]
\centering
\setCommonTableSettings
\caption{Performance metrics for Layer 2 with watermarking.}
\begin{tabular}{l c c c}
\hline
\textbf{Lambda / Beta} & \textbf{PSNR} & \textbf{MSSIM} & \textbf{BPP} \\
\hline
0.1 / 0.1 & 38.69 & 0.993 & 1.645 \\
0.05 / 0.1 & 37.57 & 0.991 & 1.211 \\
0.01 / 0.05 & 33.61 & 0.976 & 0.462 \\
0.005 / 0.02 & 31.83 & 0.964 & 0.284 \\
0.001 / 0.01 & 28.12 & 0.921 & 0.0951 \\
\hline
\end{tabular}
\label{tab:layer2WM}
\end{table}

\begin{figure}[h]
  \centering
  \includegraphics[width=1\columnwidth]{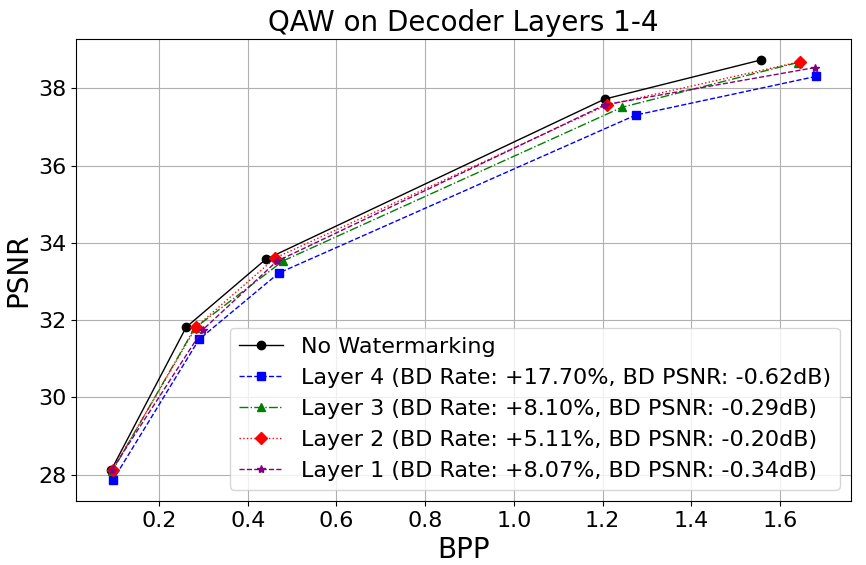}  
  \caption{RD performance of proposed QAW with pruned float model reference \cite{balle_end--end_2017} at five $\lambda$ - $\beta$ pairs}
  \label{fig:QAW_plot}
\end{figure}

\begin{table}[htbp]
\centering
\setCommonTableSettings
\caption{Performance metrics for Layer 1 with watermarking.}
\begin{tabular}{l c c c}
\hline
\textbf{Lambda / Beta} & \textbf{PSNR} & \textbf{MSSIM} & \textbf{BPP} \\
\hline
0.1 / 0.1 & 38.54 & 0.992 & 1.679 \\
0.05 / 0.1 & 37.58 & 0.990 & 1.206 \\
0.01 / 0.05 & 33.52 & 0.976 & 0.465 \\
0.005 / 0.02 & 31.74 & 0.964 & 0.298 \\
0.001 / 0.01 & 28.13 & 0.951 & 0.094 \\
\hline
\end{tabular}
\label{tab:layer1WM}
\end{table}
Our experiments corroborate this observation, as watermarking the final layer resulted in the highest performance degradation across all layers. The embedding difficulty in the final layer translates into higher $\beta$ values in the QAW loss function, reflecting a stronger penalty for watermarking during training. Consequently, this increase in $\beta$ leads to a conflict with the rate-distortion (RD) loss, ultimately resulting in a notable reduction in RD performance due to the competing objectives of watermark preservation and model accuracy.
 Furthermore, it is evident that the $\beta$ value decreases with lower $\lambda$ values. This is intuitive, as the QAW scheme jointly minimizes both loss terms, driven by the tradeoff variables $\lambda$ and $\beta$. When $\lambda$ is reduced, watermarking becomes easier, leading to an adaptive decrease in $\beta$. Finally, the QAW results are summarized in Fig.\ref{fig:QAW_plot} and Table \ref{tab:watermarking_qaware}, which further highlight the insights discussed above. Overall, Layer 2 proves to be the optimal location for watermarking, yielding the lowest RD performance loss (-0.2 dB and +5.11\% BPP) and demonstrates that the watermark is successfully recovered in nearly all cases, with only a negligible degradation in image quality compared to the float model.
 In addition, quantization further reduces the model size down to nearly 43\%, making the model more memory-efficient for deployment on resource-constrained devices. This compression is accompanied by a modest gain in inference latency (-2.56\%), improving real-time performance without substantial sacrifices in speed.

\begin{table}[h]
\centering
\setCommonTableSettings
\caption{Effect of the proposed Quantization Aware Watermarking (QAW) compared to the pruned float model}
\resizebox{\linewidth}{!}{%
\begin{tabular}{l c c c c}
\hline
\textbf{} & \textbf{Layer 1} & \textbf{Layer 2} & \textbf{Layer 3} & \textbf{Layer 4} \\ 
\hline
\textbf{$\Delta$ BD-PSNR (dB) ↑} & -0.34 & -0.20 & -0.29 & -0.62 \\ 
\textbf{$\Delta$ BD-Rate (\%)↓} & +8.07 & +5.11 & +8.10 & +17.7 \\ 
\textbf{C-BER ↑} & 97\% & 98\% & 98\% & 94\% \\  
\hline
\textbf{$\Delta$ Model size ↓} & \multicolumn{4}{c}{-42.86\%} \\
\textbf{$\Delta$ Inference latency ↓} & \multicolumn{4}{c}{-2.56\%} \\ 
\hline
\end{tabular}%
}
\label{tab:watermarking_qaware}
\end{table}

To evaluate the efficiency of QAW, we consider a simple Post Quantization Watermarking (PQW) technique where the watermark is embedded in the quantized model \cite{kakikura_collusion_2022}. In detail, a parameter tensor \( T \) from the host network is flattened, trimmed to length \( N^2 \), and reshaped into an \( N \times N \) matrix \( T_{\text{sq}} \). The watermark, generated using a key \( k \), is embedded by shuffling elements, applying the discrete cosine transform (DCT), and modifying the mid-frequency DCT coefficients. After embedding, \( T_{\text{sq}} \) is resized to match \( T \), and the host network’s parameters are updated.  
During extraction, the process is reversed: the mid-frequency domain of \( T_{\text{sq}} \) is extracted, and pseudo-random numbers are generated using \( k \). The correlation between the extracted mid-frequency domain and the generated pseudo-random numbers is calculated. A strong correlation indicates the presence of the watermark. We target the first convolutional layer and set and use the same setup variables from \cite{kakikura_collusion_2022} The model is then evaluated again on the test subset and the changes in performance are reported in Table~\ref{tab:watermarking_postq}.
While this scheme demonstrates slightly improved robustness, its adverse effect on reconstructed image/video quality is higher than that of QAW, as expected.

\begin{table}[h]
\centering
\setCommonTableSettings
\caption{Performance changes due to the reference Post-Quantization Watermarking (PQW) \cite{kakikura_collusion_2022} compared to the pruned float model
}
\begin{tabular}{l c c c}
\hline
\textbf{} & \textbf{HD} & \textbf{FHD} & \textbf{UHD} \\ 
\hline
\textbf{$\Delta$ Avg PSNR  (dB) ↑} & -0.72 & -0.93 & -0.69 \\ 
\textbf{$\Delta$ Avg MS-SSIM ↑} & -3.1\% & -7.9\% & -5.1\% \\ 
\textbf{C-BER ↑} & 99\% & 98\% & 98\% \\ 
\hline
\end{tabular}
\label{tab:watermarking_postq}
\end{table}

Our experiments demonstrated a trade-off between watermark robustness and impact on image quality. Since both methods ensure watermark presence with comparable robustness, we will only consider the QAW scheme, as it results in an average of 0.5 PSNR (dB) improvement over PQW, proving to be the better approach.

\begin{figure}[htbp]
  \centering
  \includegraphics[width=0.8\columnwidth]{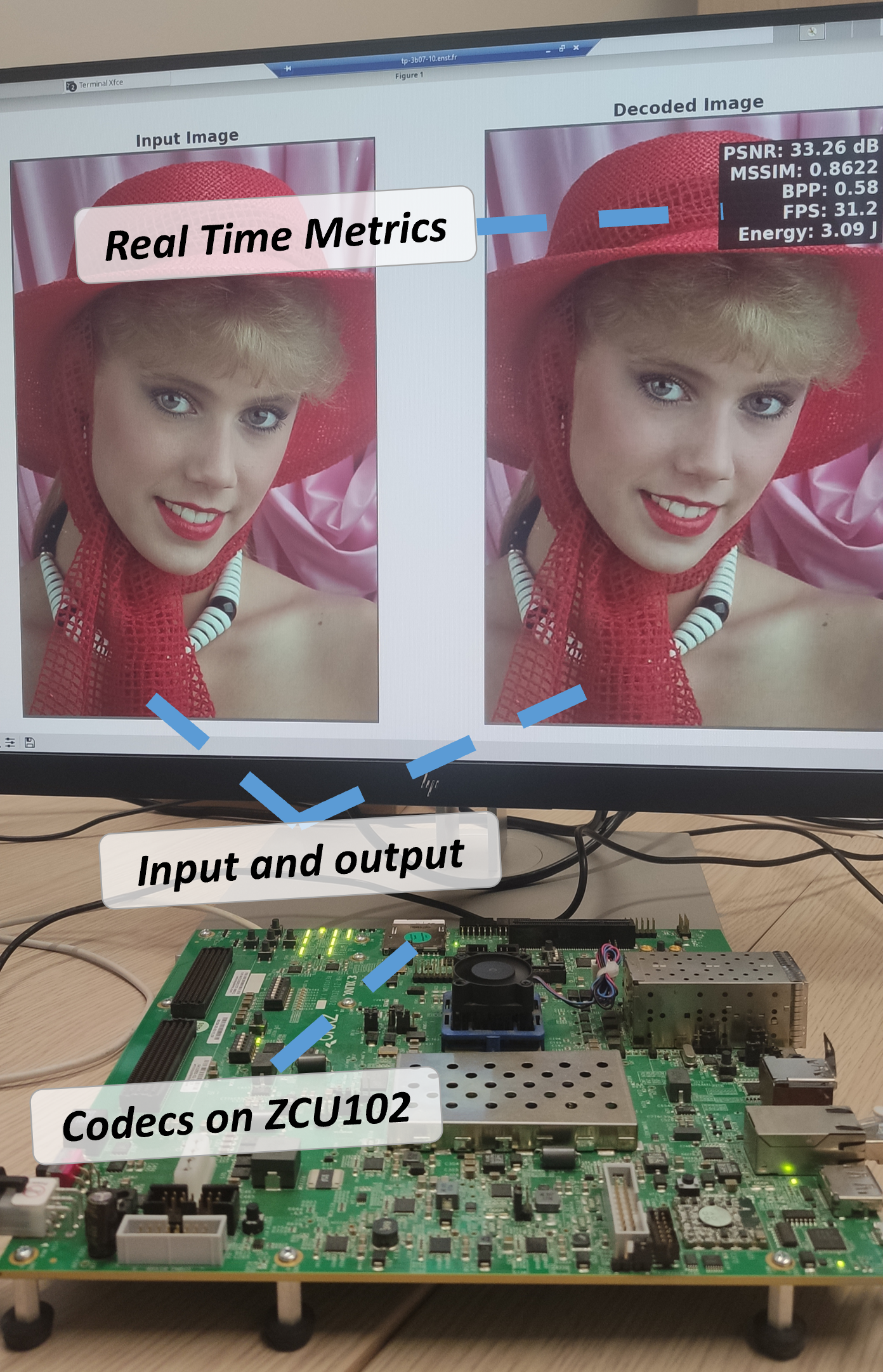}  
  \caption{ZCU102 FPGA setup during experimental validation showing real-time deployment on the embedded Linux}
  \label{fig:place_holder_fpga}
\end{figure}

\begin{table}[htbp]
\centering
\setCommonTableSettings
\caption{Example model ($\lambda$=0.01) showing changes in PSNR at different steps of the workflow evaluated on the CLIC2020 dataset}
\begin{tabular}{c c c c}
\hline
 & \textbf{HD} & \textbf{FHD} & \textbf{UHD} \\ 
\hline
\textbf{Float-Model} & 34.95 & 34.83 & 33.91 \\
\textbf{Pruning} & 33.84 & 33.15 & 33.13 \\
\textbf{QAW} & 33.61 & 32.89 & 32.71 \\
\hline
\end{tabular}
\label{tab:example_model_psnr}
\end{table}

\begin{table}[h]
\centering
\setCommonTableSettings
\caption{Power consumption comparison with related works and GPU [J/frame]}
\begin{tabular}{l c c c}
\hline
 & \textbf{HD} & \textbf{FHD} & \textbf{UHD} \\ 
\hline
\textbf{KV260 (ours)} & 3.12 & 8.59 & 31.71 \\
\textbf{ZUC102 (ours)} & 2.88 & 5.44 & 20.23 \\
\textbf{ZUC104 \cite{jia_fpx-nic_2022}} & 3.51 & 7.31 & 31.53 \\
\textbf{VCU118 \cite{sun_real-time_2022}} & 2.95 & NA & NA \\
\textbf{KU115 \cite{sun2024fpga}} & NA & NA & NA \\
\textbf{GPU Tesla K80} & 60.13 & 73.91 & 124.06 \\
\textbf{Jetson Nano (5W)} & 17.43 & 29.91 & 68.93 \\
\hline
\end{tabular}
\label{tab:power_consumption}
\end{table}

\begin{table*}[htbp]
\centering
\setCommonTableSettings
\caption{Resource utilization for HD LIC using B4096 DPU cores and comparison with state of the art}
\begin{tabular}{c c c c c}
\hline
\textbf{Clock (MHz)} & \textbf{Look-Up-Tables} & \textbf{Flip-Flops} & \textbf{BRAM} & \textbf{DSP Efficiency} \\ 
\hline
\textbf{KV260 (ours)} & 66\% & 43\% & 70\% & 774 (62\%) \\ 
\textbf{ZCU102 (ours)}  & 57\% & 54\% & 56\% & 2145 (85\%) \\ 
\textbf{VCU118 \cite{sun_real-time_2022}} & 41\% & 24\% & 34\% & 3560 (54\%) \\ 

\textbf{KU115 \cite{sun2024fpga}} & 46\% & 36\% & 28\% & 5864 (94\%) \\ 
\hline
\end{tabular}
\label{tab:resource_utilization_hd_lic}
\end{table*}

\subsection{On-board experiments}

\begin{table*}[htbp]
\centering
\setCommonTableSettings
\caption{FPS performance for the proposed hardware model using KV260 and ZCU102 FPGAs}
\begin{tabular}{l c c c c c c}
\hline
 & \multicolumn{2}{c}{\textbf{HD 1280x720}} & \multicolumn{2}{c}{\textbf{FHD 1920x1080}} & \multicolumn{2}{c}{\textbf{UHD 3840x2160}} \\ 
\cline{2-7}
 & \textbf{Hardware} & \textbf{End-to-End} & \textbf{Hardware} & \textbf{End-to-End} & \textbf{Hardware} & \textbf{End-to-End} \\
\hline
\textbf{KV260 FPS (ours)} & 26.4 & 11.9 & 9.2 & 4.8 & 4.2 & 1.38 \\
\textbf{ZCU102 FPS (ours)} & 61.2 & 36.2 & 24.2 & 13.8 & 14.3 & 3.2 \\
\hline
\textbf{ZUC104 FPS \cite{jia_fpx-nic_2022}} & \multicolumn{2}{c}{3.9} & \multicolumn{2}{c}{1.68} & \multicolumn{2}{c}{0.42} \\
\textbf{VCU118 FPS \cite{sun_real-time_2022}} & \multicolumn{2}{c}{30.28} & \multicolumn{2}{c}{NA} & \multicolumn{2}{c}{NA} \\
\textbf{KU115 FPS \cite{sun2024fpga}} & \multicolumn{2}{c}{37.74} & \multicolumn{2}{c}{20.42} & \multicolumn{2}{c}{NA} \\
\hline
\end{tabular}
\label{tab:fps_performance_fpga}
\end{table*}

\begin{figure*}[htbp]
  \centering
  \includegraphics[width=1\textwidth]{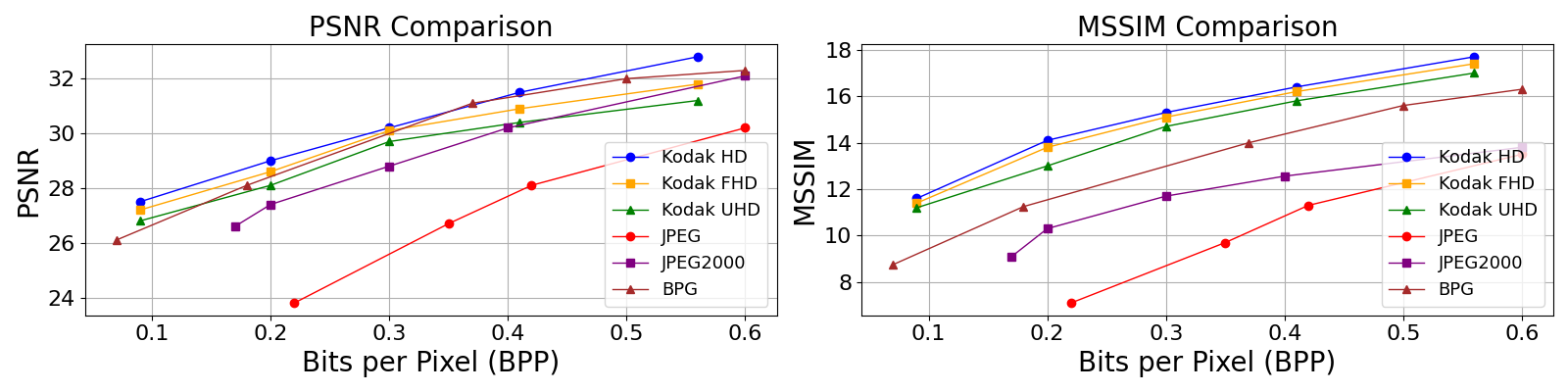}  
  \caption{PSNR (left) and MS-SSIM (right) rate curves on the Kodak dataset}
  \label{fig:kodak_plot}
\end{figure*}

\begin{figure*}[htbp]
  \centering
  \includegraphics[width=1\textwidth]{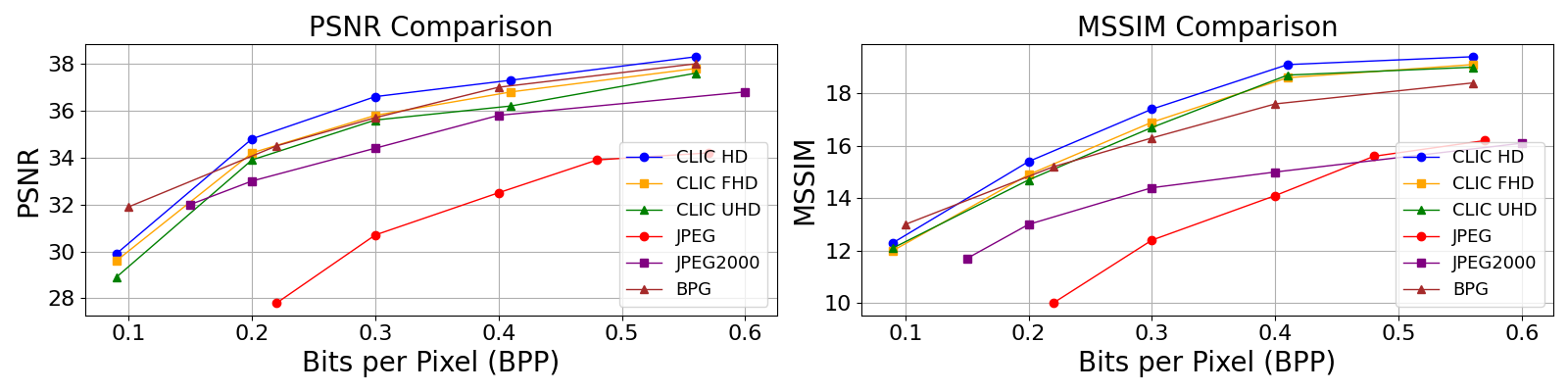}  
  \caption{PSNR (left) and MS-SSIM (right) rate curves on the CLIC2020 dataset}
  \label{fig:clic_plot}
\end{figure*}

Next, we deploy the pruned, quantized, and watermarked x-models onto the target FPGA boards for final evaluation on the ZCU102 and KV260 platforms. The hardware setup is shown in Fig.~\ref{fig:place_holder_fpga}.
The rate-distortion (RD) performance of the deployed models is analyzed on unseen images from the Kodak and CLIC datasets, as shown in Fig.\ref{fig:kodak_plot} and Fig.\ref{fig:clic_plot}, To facilitate comparison, MS-SSIM values are converted to dB, providing a more intuitive visualization of the relative quality differences between models. 

These results demonstrate that our FPGA-based compression platform achieves superior performance compared to standardized codecs across most scenarios.
Furthermore, Table~\ref{tab:example_model_psnr} quantifies the impact of the full model optimization pipeline, from the original floating-point model to the final deployed hardware implementation. 

The platform utilizes the target boards' embedded processors for pre- and post-processing, introducing software overhead, particularly in patch-based inference. This overhead is accounted for in the end-to-end latency.

\subsection{System-Level Performance and Hardware Deployment Complexity Analysis }

Table.\ref{tab:fps_performance_fpga} reports the system-level features in terms of Frames Per Second (FPS) for the hardware deployed model and the end-to-end pipeline. The combination of pruning and quantization enables significant FPS boots compared to similar works. Compared to \cite{jia_fpx-nic_2022}, using the ZCU102 which has more resources than the KV260, the proposed platform achieves a speedup of 9.2×, 8.2× and 7.6× for the HD, FHD and UHD models respectively. A slight speedup is achieved compared to \cite{sun_real-time_2022} if we assume their results are end-to-end and include software overhead. The FPS difference between UHD and FHD resolutions is primarily attributed to optimized parallelism in the FPGA's inference process. Despite UHD images containing four times the number of pixels compared to FHD, the FPGA architecture efficiently balances the increased pixel count with throughput optimization, effectively utilizing more of the available resources which results in higher DSP efficiency. Additionally, the system benefits from Vitis AI's optimized layer execution and pipelined memory access, enabling it to maintain high FPS despite the increased computational load. 

Table.\ref{tab:resource_utilization_hd_lic} shows the resource utilization for both boards and the work in \cite{jia_fpx-nic_2022}, our ZCU102 is more resource efficient as it uses 3 DPU cores for the HD LIC task. 
Table.\ref{tab:power_consumption} shows the power consumption on the ZCU102 board as measured placing shunt resistors on each of the power rails that interface with the FPGA. Each shunt resistor is linked to a TI-INA226 device, capable of measuring both voltage and current and connected to an I2C bus for convenient data collection and real-time voltage and current monitoring. We employ an embedded application running on the ARM processor to retrieve the relevant current inputs and voltage attributes, giving us PS, PL, and total power. For the KV260, we use Xilinx's font-end platform management utility—xmutil, which allows for real-time monitoring of the embedded environment.Compared to [21], our model achieves a 70\% reduction in power consumption for UHD resolution. Unlike [22], which reports results excluding software overhead, we include end-to-end performance metrics for comprehensive evaluation.
Finally, Table.\ref{tab:power_consumption} compares the proposed platform energy footprint with the related works and a GPU. Unsurprisingly, FPGAs outperform the GPU in the per-frame energy consumption; compared to \cite{jia_fpx-nic_2022} and \cite{sun_real-time_2022}, our platform exhibits a better energy budget for both end-to-end compression and decompression tasks.

\subsection{Overhead of DRM Scheme}

\begin{table*}[h]
\centering
\setCommonTableSettings
\caption{Impact of DRM on FPS and Energy (J/frame) for ZCU102. DRM refers to the combination of watermark-extraction and decryption applied to the baseline model. }
\label{tab:ZCU102_watermark_latency_fps_power}
\begin{tabular}{l c c c c c c}
\hline
\multirow{2}{*}{\textbf{Resolution}} & \multicolumn{2}{c}{\textbf{Baseline}} & \multicolumn{2}{c}{\textbf{Baseline + DRM}} & \multirow{2}{*}{\textbf{$\Delta$ Energy (\%)}} & \multirow{2}{*}{\textbf{$\Delta$ FPS (\%)}} \\
\cline{2-5}
 & \textbf{FPS} & \textbf{Power [J/frame]} & \textbf{FPS} & \textbf{Power [J/frame]} & & \\
\hline
\textbf{HD (1280x720)} & 62.5  & 2.80  & 61.2 & 2.88 & \multirow{1}{*}{+2.9\%} & \multirow{1}{*}{-2.1\%} \\
\textbf{FHD (1920x1080)} & 25.5 & 5.30  & 24.2  & 5.44  & \multirow{1}{*}{+2.6\%} & \multirow{1}{*}{-5.1\%} \\
\textbf{UHD (3840x2160)} & 15.8 & 19.80 & 14.3  & 20.23  & \multirow{1}{*}{+2.2\%} & \multirow{1}{*}{-9.5\%} \\
\hline
\end{tabular}
\end{table*}

\begin{table*}[htbp]
\centering
\setCommonTableSettings
\caption{Impact of DRM on FPS and Energy (J/frame) for KV260. DRM refers to the combination of watermark-extraction and decryption applied to the baseline model.}
\label{tab:KV260_watermark_latency_fps_power}
\begin{tabular}{l c c c c c c}
\hline
\multirow{2}{*}{\textbf{Resolution}} & \multicolumn{2}{c}{\textbf{Baseline}} & \multicolumn{2}{c}{\textbf{Baseline + DRM}} & \multirow{2}{*}{\textbf{$\Delta$ Energy (\%)}} & \multirow{2}{*}{\textbf{$\Delta$ FPS (\%)}} \\
\cline{2-5}
 & \textbf{FPS} & \textbf{Power [J/frame]} & \textbf{FPS} & \textbf{Power [J/frame]} & & \\
\hline
\textbf{HD (1280x720)} & 26.8  & 3.00  & 26.4  & 3.12  & \multirow{1}{*}{+4.0\%} & \multirow{1}{*}{-1.5\%} \\
\textbf{FHD (1920x1080)} & 9.9  & 8.40  & 9.2  & 8.59  & \multirow{1}{*}{+2.3\%} & \multirow{1}{*}{-7.1\%} \\
\textbf{UHD (3840x2160)} & 4.8  & 31.50  & 4.2  & 31.71  & \multirow{1}{*}{+0.7\%} & \multirow{1}{*}{-12.5\%} \\
\hline
\end{tabular}
\end{table*}

The results from  Tables \ref{tab:KV260_watermark_latency_fps_power} and \ref{tab:ZCU102_watermark_latency_fps_power}, demonstrate that the introduction of Quantization Aware Watermarking   and decryption (proposed DRM Scheme) into the FPGA accelerated Learned Image Compression pipeline for both KV260 and ZCU102 has a minimal impact on FPS and power consumption. For both platforms, the FPS values remain largely unaffected, with only minor reductions observed, indicating that the watermarking and decryption processes do not significantly degrade real-time performance. The small decrease in FPS, a −1.5\% for HD on KV260 and −2.1\% for HD on ZCU102, reflects the added computational overhead of DRM but does not hinder the devices' ability to handle real-time image compression tasks. Power consumption also shows a modest increase, with +4.0\% for HD and similar values for FHD and UHD, which is expected due to the additional processing required for watermark extraction and decryption. However, this increase remains within a reasonable range and should not pose a major concern in scenarios where security is prioritized. The Baseline power consumption is slightly lower than that of the Baseline + DRM scheme, as expected, but the overall power overhead is minimal. These results suggest that DRM can be effectively integrated into FPGA-based systems for secure image compression without significantly compromising performance or energy efficiency. The slight penalties in FPS and power consumption are outweighed by the enhanced security provided by DRM, making these devices suitable for applications that require secure image compression without sacrificing real-time performance.

\section{Conclusion and Discussion}
\label{sec:conclusions}

This paper presents a workflow for the real-time and secure deployment of LIC models on FPGAs. Experiments conducted on two distinct FPGA platforms demonstrate that an encoding latency of 25 FPS can be achieved for both HD and FHD content, with watermarking proving to be effectively transparent in terms of both recovered image quality and latency. 
The proposed workflow is accessible to developers with minimal hardware design expertise, enabling the training, optimization, and compilation of compression models for HD, FHD, and UHD resolutions at different distortion rates. Our experiments show that pruning and quantizing the models is the key towards reduced latency with no significant loss in RD efficiency.
Although our workflow was designed for Xilinx VITIS-AI, we believe that it can be generalized to other platforms by adapting the compilation and deployment steps while using tools such as TVM \cite{tvm} and OpenVINO \cite{intel_openvino} for final hardware compilation.
Additionally, we propose a Digital Rights Management (DRM) scheme to manage the secure usage and deployment of the hardware compression platform. This includes both encryption and watermarking. Our DRM framework offers a dual layer of security: preventing unauthorized access and enabling traceability of the model’s usage. The proposed DRM system introduces minimal throughput and energy overhead, ensuring that the compression performance remains largely unaffected while providing an essential security layer for embedded deployment. Notably, the integration of our novel Quantization Aware Watermarking (QAW) technique ensures the watermark is embedded during the quantization process, enhancing security without significant performance degradation.
The introduction of quantization aware watermarking and decryption into the FPGA-accelerated LIC pipeline results in minor FPS reductions (–1.5\%  on KV260, –2.1\% on ZCU102) and a modest power increase (+4\% on KV260 and +2.9\% on ZCU102 ). These changes remain within acceptable limits, confirming that our novel approach of embedding the watermark during quantization aware training minimizes the compression performance drop, and the DRM introducing low system-level overhead, including minimal impact on framerate and energy consumption. This ensures a secure and efficient solution for real-time applications.

\bibliographystyle{IEEEtran}  
\bibliography{LIC_2017}  

\end{document}